\documentclass[
    ,final            % use final for the camera ready runs
  ]
  {aipproc}

\layoutstyle{6x9}

\begin{document}

\title{The Planets around the Post-Common Envelope Binary NN Serpentis}


\author{Frederic V.\ Hessman}{
  address={Georg-August-Universit\"at G\"ottingen, Institut f\"ur Astrophysik, Friedrich-Hund-Platz~1, 37077~G\"ottingen, Germany}
}
\author{Klaus Beuermann}{
  address={Georg-August-Universit\"at G\"ottingen, Institut f\"ur Astrophysik, Friedrich-Hund-Platz~1, 37077~G\"ottingen, Germany}
}
\author{Stefan Dreizler}{
  address={Georg-August-Universit\"at G\"ottingen, Institut f\"ur Astrophysik, Friedrich-Hund-Platz~1, 37077~G\"ottingen, Germany}
}
\author{Tom~R.~Marsh}{
  address={Department of Physics, University of Warwick, Coventry, CV47AL, UK}
}
\author{Steven G.\ Parsons}{
  address={Department of Physics, University of Warwick, Coventry, CV47AL, UK}
}
\author{Chris M.\ Copperwheat}{
  address={Department of Physics, University of Warwick, Coventry, CV47AL, UK}
}
\author{Don~E.~Winget}{
  address={Department of Astronomy, University of Texas at Austin, RLM 16.236, Austin, TX 78712, USA}
}
\author{George F.\ Miller}{
  address={Department of Astronomy, University of Texas at Austin, RLM 16.236, Austin, TX 78712, USA}
}
\author{James J.\ Hermes}{
  address={Department of Astronomy, University of Texas at Austin, RLM 16.236, Austin, TX 78712, USA}
}
\author{Matthias R.\ Schreiber}{
  address={Departamento de F{\'\i}sica y Astronom{\'\i}a, Facultad de Ciencias, 
    Universidad de Valpara{\'\i}so, Av.~Gran Breta{\~{n}}a 1111, Valpara{\'\i}so, Chile}
}
\author{Wilhelm Kley}{
  address={Institute f\"ur Astronomie und Astrophysik, 
    Kepler Center for Astro and Particle Physics, 
    Eberhard-Karls-Universit\"at T\"ubingen, 
    Morgenstelle 10, 72076 T\"ubingen, Germany}
}
\author{Vik S. Dhillon}{
  address={Department of Physics and Astronomy, University of Sheffield, Sheffield S3 7RH, UK}
}
\author{Stuart P. Littlefair}{
  address={Department of Physics and Astronomy, University of Sheffield, Sheffield S3 7RH, UK}
}

\begin{abstract}
We have detected 2 circumbinary planets around the close binary system NN Serpentis using the orbital time delay effect measured via the sharp eclipses of the white dwarf primary.     The present pre-cataclysmic binary was formed when the original $\sim 2\,M_\odot$ primary expanded into a red giant, causing the secondary star to drop from its original orbit at a separation of about $1.4\,$A.U. down to its current separation at $0.0043\,$A.U.    A quasi-adiabatic evolution of the circumbinary planets' orbits during the common-envelope phase would  have placed them in unstable configurations, suggesting that they may have suffered significant orbital drag effects and were originally in much larger orbits.   Alternatively, they may have been created as 2nd-generation planets during the last million years from the substantial amount of material lost during the creation of the  binary, making them the youngest planets known.  Either solution shows how little we actually understand about planetary formation.
 \end{abstract}

\maketitle

%%%%%%%%%%%%%%%%%%%%%%%%%%%%%%%%%%%%%%%%%%%%
%% MAINMATTER
%%%%%%%%%%%%%%%%%%%%%%%%%%%%%%%%%%%%%%%%%%%%

\section{Introduction}

NN\,Ser is a detached close binary system consisting of a dwarf M star in a very close orbit ($P_{\rm orb}=3.1\,$hr) around the hot white dwarf primary.
Such close binaries are formed when the separation of the original main sequence stars are smaller than the extent of a red giant: when the envelope of the evolved primary expands past the orbit of the secondary, frictional forces transfer the orbital angular momentum and energy from the secondary to the envelope.
During the resulting ``common-envelope'' (CE) phase, the secondary plunges down into the final tight orbit and $3/4$ of the primary's mass is cast off in the form of a planetary nebula.
With an unusually high temperature of about $50,000\,$K for the white dwarf in NN\,Ser, this event must have happened only about a million years ago.
Eventually, the loss of angular momentum due to stellar winds and gravitational radiation will bring the white dwarf and M star into contact, after which NN\,Ser will become a classic cataclysmic variable (CV).

The light curves of eclipsing pre-CV's are particular simple, since the light curves are affected only by the varying visibilities of the highly irradiated dwarf and the dramatic eclipses of the white dwarf.
This fact, along with the very short orbital periods, makes it particularly easy to measure the ephemerides of such systems.
For several of these systems, long-term monitoring has resulted in the detection of highly significant orbital period variations of the order of tens of seconds relative to linear long-term mean ephemerides (Parsons et al.\ 2010b).
There are various plausible origins for these variations, including Applegate's mechanism (Applegate 1992), apsidal motion or the presence of one or more bodies around the binary.
Historically, there has been some hesitance about attributing the observed timing effects to one or more circumbinary objects: not only does one have to exclude other possible explanations, there is also a deep-set skepticism about the idea of the formation of planets around binary star systems.
Thus, the search for circumbinary planets is interesting on several accounts: how many are out there, what binary systems can support them, and even when are such planets formed?
Post-CE systems like NN\,Ser provide a unique opportunity to answer some of these questions.

\section{Re-analysis of previous observations}

The timing data in the literature for NN\,Ser consisted of the early photoelectric measurements by Haefner et al.\ (1989) on the ESO Danish 1.5m,  CCD photometry on a smaller telescope by Pigulski \& Michalska (2002), VLT observations by Haefner et al.\ (2004), WHT+UltraCam observations by Brinkworth et al.\ (2006), and finally CCD photometry on a $2.4\,$m by Qian et al.\ (2009).
The latter group looked at all of the very heterogeneous timing measurements and suggested the deviations might be due to a 3rd body -- a massive Jupiter-like planet in a $7\,$year orbit -- but they ignored the fact that some of the measurements were very precise and did not fit their simple 3-rd body orbit at all ($\sim 100\!-\!\sigma$ deviations).
Part of the problem was the fact that there were too few early measurements with accuracies comparable with the latter measurements, particularly an inexact measurement by Haefner et al.\ (2004) using a trailed image and the VLT, for which Haefner et al.\ estimated an accuracy of only $17\,$sec.

In the hopes of extracting a more accurate timing measurement from the 1999 VLT observations, we re-analyzed the trailed images of NN\,Ser taken by Haefner et al.\ (2004).
Only the image taken on 11 June 1999 was good enough for a detailed analysis (Fig.\,1).
There are several problems associated with the time-calibration of trailed images, including the intrinsic accuracy of the system clock and potential absolute delays caused by camera operations as well as the relative timing accuracy posed by the characterization of the intrinsic length of the trail.
The absolute precision of the times were estimated by ESO staff to be better than $10\,$ms.
The measurement of the length of the trail was performed using several methods.
Using star images in acquisition images taken immediately before and after the trailed image, it was possible to estimate the intrinsic smearing of the trail along the trail axis, from which the starting and ending points could be determined; together with the integration time, this determined the time-scaling.
Alternatively, one can simply take the known plate scale and integration time and calculate the time scaling, checking only for a comparable seeing conditions at the start and end of the exposure.
Using the Lucy deconvolution algorithm and the acquisition images, the trails of neighboring stars were deconvolved to produce trails with very sharp boundaries, which then could be compared with the NN\,Ser trail.
Finally, the length of the eclipse in the extracted pixel light curve can be calibrated to the known shape of the eclipse, producing a nearly independent time scaling.
All of these methods led to a very exact common time calibration and a final central eclipse time of BJD(TT) $2451340.7165402(23)$, i.e. with an accuracy of about $0.2\,$sec.

In addition, we re-analyzed the CCD photometric data of Pigulski \& Michalska (2002), kindly provided in the form of a light curve by the authors.
Although they had fit a formal eclipse model profile to the data, the finite integration times had not been taken into account.
Since the eclipse is of known width and almost exactly representable by a trapezoid of known width and central depth, it was easy to correct for this effect, resulting in a significant improvement in this timing point as well (from $34.6$ to $8.3\,$sec): BJD(TT) $2451667.478006(96)$.

%%%%%%%%%%%%%%%%%%%%%%%%%%%%%%%%%%%%%%%%%%%%
\begin{figure}
  \includegraphics[width=.6\textheight]{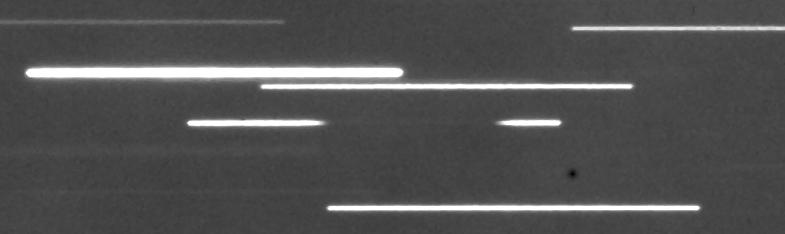}
  \caption{Trailed VLT/FORS image of NN\,Ser taken by Haefner et al.\ (2004) on 1999-JUN-11.}
\end{figure}
%%%%%%%%%%%%%%%%%%%%%%%%%%%%%%%%%%%%%%%%%%%%

\section{New observations}

We have obtained additional observations using several telescopes.
A small number of highly accurate timings was obtained with UltraCam on the ESO 3.5m NNT telescope with a Sloan g' filter (orbits $E=61219$ \& $61579$) and with the Argus camera on the McDonald Observatory $2.1\,$m with a BG40 (six eclipses between $E=60927$ and $61564$).
These measurements were supplemented with white-light observations made with the remotely controlled the MONET/North telescope at McDonald (seven eclipses between $E=60489$ and $60774$).   
The resulting table of revised and new timing observations can be found in Beuermann et al.\ (2010).

The timing points from the literature, the revised timing values, and our new observations are shown in Fig.\,2.
With the revised timing points, the variations no longer look random but show a quasi- but not perfectly sinusoidal change in the effective orbital period on a timescale of two decades.

\section{Analysis}

In the case of NN\,Ser, it is very easy to see that the timing variations can only be due to at least one additional body.
The observed timing variations are much too large to be due to the Applegate mechanism (Chen 2009).
While apsidal motion is possible, the amplitude and form depend upon the eccentricity of the central binary orbit: since the observed eccentricity is essentially zero (Parsons et al.\ 2010a), the apsidal effect should be small and sinusoidal.  

We first attempted a one-planet fit ($7$ free parameters, including the binary ephemeris) to the timing variations.
The resulting orbit has a large eccentricity ($>0.65$) and an orbital period of $22.6\,$years.
However, the reduced $\chi^2$ is $22.88$, showing that the residuals on shorter timescales are highly significant.
A full two-planet fit ($12$ free parameters) suffers from the bunching of the data points into $14$ loose groups, so a full mapping of parameter space was necessary in order to find a reasonable family of solutions.
Since the eccentricity of the main planetary component is always small, we set it to zero and fit the remaining $10$ parameters.
Two solutions remained, both having reduced $\chi^2$ of about $0.8$: one with an orbital period ratio of $2:1$ and another with a ratio of $5:2$, suggesting that the two planets are in a resonant configuration.
The planet with the largest effect has an orbital period of $15.5$ (or $16.7$) years, a semi-major axis of $5.4$ ($5.7$) A.U., and a mass of $6.9$ ($5.9$) $M_{\rm Jupiter}$ (assuming that the planets have the same inclination of $89^\circ$ as the binary).
The companion planet has a shorter period of $7.8$ ($6.7$) years, a semi-major axis of $3.4$ ($3.1$) A.U., and a mass of $2.2$ ($1.6$) $M_{\rm Jupiter}$.
The latter also has a significant eccentricity of $0.20$ ($0.23$), so that the second timing component cannot be due to apsidal motion but must be due to a true additional body.
The detailed parameters and errors can be found in Beuermann et al.\ (2010).

NN\,Ser thus joins the very small group of binary star systems with circumbinary planets: DP\,Leo (one planet; Qian et al.\ 2010), HU\,Aqr (one planet; Schwarz et al.\ 2009), QS\,Vir (probably one planet in addition to a brown dwarf; Parsons et al.\ 2010b), and  HW\,Vir (2 planets; Lee et al.\ 2009).

%%%%%%%%%%%%%%%%%%%%%%%%%%%%%%%%%%%%%%%%%%%%
\begin{figure}
  \includegraphics[height=.6\textheight]{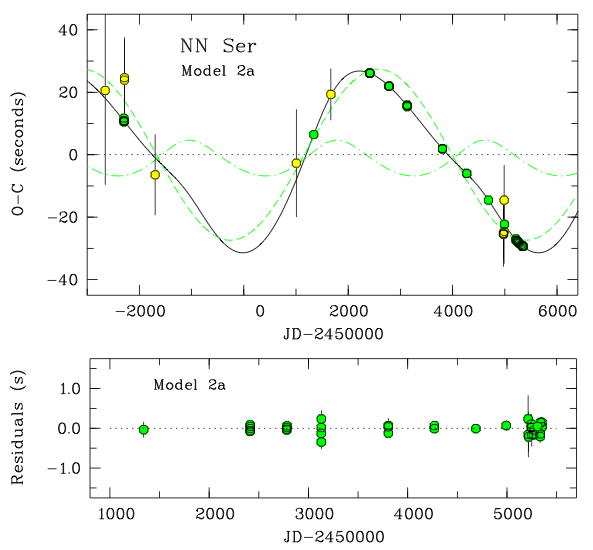}
  \caption{Eclipse time variations of NN\,Ser.  Top: observed minus calculated eclipse timings relative to a linear binary ephemeris; the errors for the green points are smaller than the symbols and the green dashed lines are the individual timing curves for each planet.  Bottom: residuals from the two-planet fit for the time period of the new observations, showing residuals consistent with the measurement errors.}
\end{figure}
%%%%%%%%%%%%%%%%%%%%%%%%%%%%%%%%%%%%%%%%%%%%

\section{Discussion}

In order to characterize the planetary system in NN\,Ser, one needs to know how, where, and when the planets were formed.
Unfortunately, it is not at all trivial to estimate what the binary system looked like before the close binary was created: the CE phase is very difficult to simulate.
The frictional cross-sections of the secondary star and planets are largely gravitational, for which the Bondi-Hoyle drag coefficients are only a rough approximation.
The secondary star plunging into the envelope of the red giant primary is not simply a question of a transfer of energy and angular momentum because the envelope as a quasi-hydrostatic body has it's own internal energy.
The CE itself assumes a strongly bipolar shape as it expands but the actual density and velocity of the envelope in the orbital plane -- needed to calculate the drag effects  -- is difficult to estimate.
In the late stages, the hot core of the red giant creates an enormous bipolar H\,{\sc II} region and thus dramatically increases the inner sound speed relative to the cool expanding outer envelope.
Zorotovic et al.\ (2010) have studied the simple parameterizations of CE and found that the CE efficiency parameter, $\alpha_{\rm CE}$, must be in the range $0.2$--$0.3$.
Adopting this parameterization and a value of $\alpha_{\rm CE}=0.25$, we estimate that the former binary star system consisted of a $\sim 2\,M_\odot$ A star orbited by the present M4 dwarf secondary at a separation of about $1.44\,$A.U.

One possibility is that the planets in NN\,Ser existed before the CE phase and were created with the stars out of an original circumbinary proto-planetary disk -- so-called ``1st-generation'' planets.
Stability requires that such planets have orbits with semi-major axes of at least $3.5\,$A.U.
Due to the loss of $3/4$ of the mass of the binary system during the CE, one would naively expect any 1st-generation planets to either be lost or at least to have their orbits expanded quasi-adiabatically by a factor of $\sim \sqrt{2.0/0.6} \approx 2$.
Thus, 1st-generation planets must have originally been at much larger orbital separations and have suffered the orbital drag during the CE phase which produced the close binary, even though the gravitational cross-section is much smaller and hence the intrinsic effect less strong.
If drag is more important than one would naively expect, does this mean that the original planets may have been formed in very distant orbits and how many close-in planets were lost?

The alternative scenario is that the planets did not exist before the CE phase but were created afterwards.
This is not implausible, since the amount of matter cast off is substantial -- much more than is typically in a proto-planetary disk.
However, it is not at all clear how the planetary nebula caused by the CE evolves over long timescales: if the expansion of the nebula is sub-sonic, does this mean the matter ``returns'', forms a circumbinary disk, and creates 2nd-generation planets?
Does the enriched composition of the ejecta make it simpler to form planets?
Is it possible to turn 1st-generation small planets/asteroids into massive 2nd-generation Jupiters?
Given the maximum age of 1 million years, 2nd-generation planets in NN\,Ser would be the youngest known by far, making them planets particularly interesting for a wide variety of reasons.

\section{Conclusions}

The recognition that NN\,Ser has a complex circumbinary planetary system with an enigmatic history and that there are bound to be many more of these systems out there opens up a totally new perspective on the formation of planets.
Before these systems appeared, we naively thought it was very unlikely that circumbinary planets might form.
Now we know that they exist, but realize how little we actually know about their cosmogony.

%%%%%%%%%%%%%%%%%%%%%%%%%%%%%%%%%%%%%%%%%%%%%%%%
%% BACKMATTER
%%%%%%%%%%%%%%%%%%%%%%%%%%%%%%%%%%%%%%%%%%%%%%%%

%\begin{theacknowledgments}
%\end{theacknowledgments}



\bibliographystyle{aipproc}   % if natbib is available

% \bibliography{sample}
% \end{document}

%%%%%%%%%%%%%%%%%%%%%%%%%%%%%%%%%%%%%%%%%%%
%% The following lines show an example how to produce a bibliography
%% without the help of the BibTeX program. This could be used instead
%% of the above.
%%%%%%%%%%%%%%%%%%%%%%%%%%%%%%%%%%%%%%%%%%%

\end{document}